%
%

\input harvmac.tex

\def\IR{\relax{\rm I\kern-.18em R}}
\def\IZ{\relax\ifmmode\mathchoice
{\hbox{\cmss Z\kern-.4em Z}}{\hbox{\cmss Z\kern-.4em Z}}
{\lower.9pt\hbox{\cmsss Z\kern-.4em Z}}
{\lower1.2pt\hbox{\cmsss Z\kern-.4em Z}}\else{\cmss Z\kern-.4em
Z}\fi}
\font\cmss=cmss10 \font\cmsss=cmss10 at 7pt

\def\pl#1#2#3{Phys. Lett. {\bf #1B} (#2) #3}
\def\prl#1#2#3{Phys. Rev. Lett. {\bf #1} (#2) #3}
\def\physrev#1#2#3{Phys. Rev. {\bf D#1} (#2) #3}

\def\ptp#1#2#3{Prog. Theor. Phys. {\bf #1} (#2) #3}
%


\lref\cds{A. Connes, M. R. Douglas and A. Schwarz, ``Noncommutative
Geometry and Matrix Theory: Compactification on Tori,'' hep-th/9711162.}%

\lref\natinf{N. Seiberg, ``New Theories in Six Dimensions and Matrix 
Description of M-theory on $T^5$ and $T^5/Z_2$'', hep-th/9705221}%

\lref\edhiggs{E. Witten, ``On the Conformal Field Theory of the Higgs Branch'',
hep-th/9707093}

\lref\ednew{E. Witten, ``New 'Gauge' Theories in Six Dimensions'', 
hep-th/9710065}%

\lref\savfive{S. Sethi, ``The Matrix Formulation of Type IIB Five-Branes, 
hep-th/9710005}%

\lref\dvv{R. Dijkgraaf, E. Verlinde and H. Verlinde, 
``5-D Black Holes and Matrix Strings'', hep-th/9704018, 
Nucl. Phys. B506, 121-142, 1997}%

\lref\dh{M. Douglas and C. Hull, ``D-Branes and Non-commutative Geometry'',
hep-th/9711165}%

\lref\bigbook{A. Connes, ``Noncommutative Geometry'', Academic Press, 1994}

\lref\abkss{O. Aharony, M. Berkooz, S. Kachru, N. Seiberg, and E. 
Silverstein, 
``Matrix Description of Interacting Theories in Six Dimensions,'' 
hep-th/9707079.}%

\lref\abs{O. Aharony, M. Berkooz and N. seiberg, ``Light-Cone Description 
of (2,0) Field Theories in Six Dimensions'', hep-th/9712117}%

\lref\os{O. Ganor and S. Sethi, ``New Perspectives on Yang-Mills Theories 
with Sixteen Supersymmetries'', hep-th/9712071}

\lref\nakajima{H. Nakajima, ``Resolutions of Moduli Spaces of Ideal 
Instantons on $\IR^4$,'' in ``Topology, Geometry and Field Theory,''
ed. Fukaya, Furuta, Kohno and Kotschick, World Scientific.}

\lref\ns{N. Nekrasov and A. Schwarz, In preparation}

\lref\abks{O. Aharony, M. Berkooz, S. Kachru, and E. 
Silverstein, 
``Matrix Description of $(1,0)$ Theories in Six Dimensions,'' 
hep-th/9709118.\semi D. A. Lowe, ``$E_8\times E_8$ Small Instantons in Matrix
Theory,'' hep-th/9709015.}%

\lref\old{A. Casher, ``Gauge Fields on the Null Plane,'' 
\physrev{14}{1976}{452}\semi 
C. B. Thorn, ``Quark Confinement in the Infinite Momentum
Frame,'' \physrev{19}{1979}{639}\semi 
C. B. Thorn, ``A Fock Space
Description of the $1/N_c$ Expansion of Quantum Chromodynamics,''
\physrev{20}{1979}{1435}\semi 
C. B. Thorn, ``Asymptotic Freedom in the
Infinite Momentum Frame,''
\physrev{20}{1979}{1934}\semi
T. Maskawa and K. Yamawaki, ``The Problem of $P_+=0$ Mode
in the Null Plane Field Theory and Dirac's Method of Quantization,''
\ptp{56}{1976}{270}\semi
H. C. Pauli and S. J. Brodsky, ``Solving Field Theory in
One Space One Time Dimension,'' \physrev{32}{1985}{1993}\semi
H. C. Pauli and S. J. Brodsky, ``Discretized Light Cone Quantization :
Solution to a Field Theory in One Space One Time Dimensions,''
\physrev{32}{1985}{2001}.}%

\lref\natidlcq{N. Seiberg, ``Why is the Matrix Model Correct ?,''
\prl{79}{1997}{3577}, hep-th/9710009.}

\lref\berdoug{M. Berkooz and M. Douglas, ``Five-branes in M(atrix) 
Theory,''
\pl{395}{1997}{196}, hep-th/9610236.}%

\lref\adhm{M. Atiyah, V. Drinfeld, N. Hitchin, and Y. Manin,
``Construction of Instantons,'' \pl{65}{1978}{185}.}

\lref\bfss{T. Banks, W. Fischler, S. Shenker, and L. Susskind, 
``M theory as a Matrix Model:  A Conjecture,'' hep-th/9610043,
\physrev{55}{1997}{112}.}%

\lref\susskind{L. Susskind, ``Another Conjecture About Matrix Theory,''
hep-th/9704080.}%

\lref\twozero{E. Witten, ``Some Comments on String Dynamics'', 
hep-th/9507121, Contributed to Strings 95; A. Strominger ``Open P-branes'', 
hep-th/9512059, Phys. Lett. 383B (1996) 44.}%




\Title{\vbox{\baselineskip12pt\hbox{hep-th/9802069}
\hbox{IASSNS-HEP-98/12, NSF-ITP-98-018}}}
{\vbox{\centerline{Non-local Field Theories and the Non-commutative
Torus}}}

\centerline{Micha Berkooz} 
\smallskip
\smallskip
\centerline{Institute for Advanced Study}
\centerline{Princeton, NJ 08540.}
\centerline{and}
\centerline{Institute for Theoretical Physics}
\centerline{University of California}
\centerline{Santa-Barbara, CA, 93106.}
\centerline{\tt berkooz@ias.edu}
\bigskip
\bigskip
\noindent

We argue that by taking a limit of SYM on a non-commutative torus one
can obtain a theory on non-compact space with a finite non-locality
scale. We also suggest that one can also obtain a similar
generalization of the (2,0) field theory in 5+1 dimensions, and that
the DLCQ of this theory is known.

\Date{February 1998}


\newsec{Introduction}

Among the spin-offs of Matrix theory \bfss\ are new kinds, or new
formulations, of field theories (or other theories that may not have
standard gravity in them
\refs{\natinf,\ednew,\abkss,\edhiggs,\savfive,\abks,\dvv,\os}). An interesting
class of such theories arises in the Matrix description of M-theory
compactified on a torus with a non-zero 3-form Wilson line
\refs{\cds,\dh}. The resulting theories on the world-volume of
the Matrix description are Super-Yang-Mills on a non-commuting
manifold \bigbook. 

Beyond their application for Matrix theory, these theories are
interesting in their own right. One reason is that one loses the
notion of exact locality of the theory and it only reappears, in some
cases, as an approximate notion for the low-energy (or low
wave-length) observer. In particular, the theory is no longer scale
invariant in the UV. This is in contrast with more conventional field
theories. One of the elements that make ``standard'' field theory work
is the existence of a UV fixed point. Using this fixed point one
obtains a theory in the continuum. The theory on the non-commuting
torus seems to be defined in the continuum by some other
means. Returning to the low energy observer, such an observer may see
a renormalization group trajectory, if such a notion is still useful,
coming from an unexpected region of field theory parameter space.
 
In the following we discuss some of the above statements. We argue
that it is possible to disentangle the UV from the IR by taking a
limit such that the size of the torus goes to infinity, keeping the size of
the non-locality fixed. This configuration, as the one in \dh\ also
arises within string theory. To better understand the UV of the theory
we note that by a similar procedure we can obtain a non-local version
of the (2,0) field theory \twozero. A Matrix description of the latter theory
exists and may be more useful in the study of the UV of these
theories.
   
As this work was completed we were informed that N. Nekrasov and
A. Schwarz have obtained related results \ns.

\newsec{A Limit of Super-Yang-Mills on the Non-commutative 2-Torus}

Our goal in this section is to obtain a deformation of SYM on
2+1 non-compact dimensions such that it is characterized by a finite
non-locality length. Operationally, we will be interested in a
deformation of the SYM theory of the form (x,y,z being appropriate
3-vectors)
\eqn\dfrm{\int_{T^2\times R} d^3x\phi_1(x)
e^{i\zeta {\partial\over\partial y}{\partial\over \partial z}} 
\phi_2(y)\phi_3(z)\vert_{y=z=x}} where $\phi_i$ are some fields in the SYM. 
The requirement of a finite non-locality length is that as we take the
size of the spatial torus to infinity, $\zeta$ (of mass dimension
-2) remains finite. It then characterizes a finite non-locality length
in the theory on $R^2$. A low-energy observer (below the scale set by
$\zeta$) may consider the expansion in $\zeta$ as an expansion in
momenta of some effective action.

To obtain such a theory, one uses a variant of the procedure used by
Douglas and Hull. Throughout the discussion we will set $\alpha'=1$.
These authors considered a compactification of the IIA string on a
torus with
$$G+B={\Bigl(\matrix{R_1^2&B\cr -B&R_2^2}\Bigr)}$$ as
$R_1=R_2\rightarrow0$ and $B$ held fixed. We will be interested in
\eqn\varianta{G+B=\Bigl(\matrix{R_1^2&\eta R_1R_2\cr -\eta R_1R_2&R_2^2}\Bigr)}
in the limit in which $R_1=R_2=R\rightarrow 0$ and $\eta$ is held
fixed.  In this limit one keeps the value of the B field fixed (in
physical coordinates, where $G=I_{2\times 2}$) rather than the flux of
the B field through the 2-torus.

When we now perform a T-duality in both directions we
obtain that the new $G+B$ is
\eqn\invrtgb{G+B={1\over R_1^2R_2^2(1+\eta^2)}
\Bigl(\matrix{R_2^2&-\eta R_1R_2\cr \eta R_1R_2&R_1^2}\Bigr).}
The T-dual radii go to infinity but so does the $\int_{T^2} B$. In
particular, we need to perform different $SL(2,Z)$ transformations in
order to map this configuration into the fundamental domain. These
SL(2,Z) transformations are weaker than the ones in \dh\ as they
involve only shifts in $\int B$. The trajectory is therefore not
ergodic. Still, the $SL(2,Z)$ transformations that are needed are
rapidly varying when $R\rightarrow 0$, and it is not clear how to take
the limit of the entire string theory. 

A somewhat clearer picture appears if we continue with the rest of the
analysis in \dh\ (T-dualizing a single circle etc.), adapted to this
case. Inserting the current value of B into the action there (equation
(2)) one obtains a non-local term of the form,
$${S_{int}\propto\int_{0<\sigma^i<1} d\sigma^1d\sigma^2 
\phi_1(\sigma^1,\sigma^2)
exp{\bigl({\eta R_1R_2\over 2\pi i}
{\partial\over\partial \sigma^{1'}}
{\partial\over\partial \sigma^{2'}}\bigr)}
\phi_2(\sigma^1,\sigma^{2'})\phi^3(\sigma^{1'},\sigma^2)\vert_{\sigma^{i'}=
\sigma^i}}$$

We can now go back to the physical coordinates $x^i={\sigma^i\over
R_i}$ on the dual torus and obtain a finite non-locality size, in the
form of \dfrm, while the size of the torus is taken to infinity.

\newsec{Relation to the Non-local (2,0) field theory}

\subsec{The Limit Procedure for the D4-brane}

Let us repeat the same procedure as before but for the D4-brane. For
convenience let us divide the 4-torus into a product of 2-tori and turn
on a B field separately in each of these. We can now copy the result of
the previous section, and obtain a 4-brane on a torus of radii
$R_1,..,R_4$ which go to infinity and a finite B field, inducing a
finite non-locality length. We will again parameterize the non-locality
by a parameter of $\zeta$ of mass dimension $-2$. $\zeta$ is related to
$B$ (in the frame where $G=I_{4\times 4}$) by
$$B={\zeta\over {\alpha'}^2}.$$

We are interested in lifting this configuration to M-theory. The
background $B$ field now becomes a background $C_3$ field, given by
$$B=R_{11}C,$$ and the D4-brane is now to be interpreted as a wrapped
M5-brane. As discussed in \abs, even though the $C$ field is constant,
it is not pure gauge. This is so because of the M5-brane. There is now
a gauge invariant quantity, $H-C$, which is not zero in our
case. Phrasing it in another way, we can gauge away $C$ and generate a
non-zero $H$ field. This is a field strength and is not pure
gauge. Using the above relations one finally obtains
\eqn\finc{\zeta={C\over R_{11}M_p^6}.} 
A DLCQ of a non-local deformation of the (2,0) field theory with
exactly this non-locality scale was proposed in \abs\foot{The choice
of C there was actually more restrictive. This will be discussed
later.}. One therefore conjectured that the non-local deformation of
the (2,0) theory in 5+1 dimensions in non-compact space is a
well-defined theory. There is also a concrete quantum mechanical
system that describes the DLCQ of this theory.

Upon compactification, this theory becomes the natural definition of
the non-local deformation of 4+1 SYM (or 4+1 SYM on the
noncommutative torus). By appropriately compactifying the theory one
expects to obtain similar non-local SYM on lower dimensional $R^d$. It
may be that deformations of the models described in \os\ are Matrix
models of these SYM.

In the following subsection we will review some facts about the Matrix
description of the non-local deformation of the (2,0) field
theory. The discussion will include only basic facts that will be
useful for the a subsequent brief discussion. For a more complete
picture the reader is referred to \abs.

\subsec{The Non-local (2,0) field theory}

Part of the analysis in this subsection, as well as part of the
following section, is extracted from \abs. The reader interested in
more details and in a more complete picture is referred there. It is
brought here for completeness, and in order to set up some details
that will be necessary in the discussion later.

The Matrix description of the DLCQ \refs{\old,\susskind,\natidlcq} of the
$(2,0)_k$ field theory, at $P_-$ momentum sector $N$, is given by quantum
mechanics on the moduli space of N-instantons in $U(k)$ on $R^4$
\refs{\abkss}. This manifold is described in the ADHM \adhm\ construction by
the following HyperK\"ahler quotient.

The initial space is a linear representation of $U(N)$. It is given by
2 complex N*N matrices, $X$ and $\tilde X$, and by two complex N*K
matrices $Q$ and $\tilde Q$. This space admits an action of $U(N)$ by
\eqn\rgpact{X\rightarrow gXg^{-1},\ {\tilde X}\rightarrow g{\tilde X}g^{-1},\ 
Q\rightarrow gQ,\ {\tilde Q}\rightarrow g^{-1}{\tilde Q}.}
The hyperK\"ahler quotient is then given by the $X,{\tilde X},Q,{\tilde
Q}$ 
\eqn\hypklr{\Biggl(\matrix{
[X,X^\dagger]-[{\tilde X},{\tilde X}^\dagger]+QQ^\dagger-{\tilde
Q}^*{\tilde Q}^T=0\cr\cr [X,{\tilde X}]+Q{\tilde Q}^T=0}\Biggr)/U(N)}

Locality in spacetime manifests itself in the quantum mechanics in the
following way. The theory in spacetime is in a superconformal fixed
point. When we go to the DLCQ the light-like circle breaks most of
the conformal invariance. It leaves however an SO(1,2) subgroup,
which becomes the conformal symmetry of the quantum mechanics. One of
the generators of the latter, which we will denote by $T$, can be
identified as $T=D+M_{01}$, where $D$ is the dilatation operator in
spacetime and $M_{01}$ is the generator of Lorentz transformations in
the $X^+-X^-$ plane. $T$ acts in the quantum mechanics by contracting
the entire sigma model to the origin.

One way to see that the theory is local is the following. Take a state
in the quantum mechanics, i.e. some normalizeable wave function on the
quantum mechanics, and act on it by $T$. The support of the wave
function shrinks to the origin. As $T$ contains $D$ in it, the
spacetime interpretation of this operation is that we take some state
in spacetime and shrink it to the point. We therefore obtain a local
state, or local operator, in spacetime. The fact that we were not
obstructed in this process implies that the theory has local objects
in it.

The analysis in \abs\ included a resolution of the singularities in
the space. The resolved space \nakajima\ is given by
\eqn\rslv{\Biggl(\matrix{
[X,X^\dagger]-[{\tilde X},{\tilde X}^\dagger]+QQ^\dagger-
{\tilde Q}^*{\tilde Q}^T=\zeta I_{N\times N}\cr\cr
[X,{\tilde X}]+Q{\tilde Q}^T=0}\Biggr)/U(N).} 
  
In \abs, this resolution was used primarily to analyze the $(2,0)_k$
field theory, but it was also noted that the theory is interesting in
its own right. The theory is now non-local and does not posses scale
invariance. Now we cannot shrink the support of the function to the
origin, since the origin is no longer in the manifold. It has been
replaced by some manifold of size $\zeta$. On scales larger the
$\zeta$ we can approximately shrink the support of the function but
this procedure fails when we reach the scale set by $\zeta$. The
implications in spacetime are that on larger length scales the theory
will appear as a deformation of a local theory, but we will begin to
feel the non-locality of the full theory at a scale set by
$\zeta$. This is exactly the situation that we have for SYM on a
noncommuting manifold.

Furthermore, thinking about the deformed $(2,0)_k$ theory as the
theory on the M5-brane, it is shown in \abs\ that the deformation is
given by turning on a certain C field, and that the scale of
non-locality is given by
\eqn\czeta{\zeta={C_{+ij}\over RM_p^6}}
which is the same relation that we had before.

One more issue remains to be discussed. The world-volume of the
D4-brane has an $SO(4)\sim SU(2)_R\times SU(2)_L$ symmetry of spatial
rotation. In the previous subsection, it seemed that we were allowed
to turn on a background B field with arbitrary SO(4) chirality. That
is not the case with the $\zeta$ parameter in \abs. $SU(2)_R$ is an R
symmetry of the QM which rotates the different constraint equations in
the ADHM construction and $\zeta$ takes value in the $(3,1)$ of
$SO(4)$. The reason for the restriction is not a fundamental one in
the theory, rather it is an artifact of the DLCQ description. We have
chosen $P_->0$ momenta in a certain lightlike direction and we define
what we call 0-branes accordingly. In the sector with $P_->0$, turning
on the $H$ field with the wrong $SO(4)$ chirality will break
supersymmetry. We have, however, restricted ourselves to deformation
that preserve supersymmetry, and therefore we can not describe turning
on all the $SO(4)$ chiralities of $C$.

\newsec{A Comment on the UV of the Non-local Theories}

Some brief comments are due:

\noindent 1. The fact that we have obtained a Matrix description of the 
non-local $(2,0)_k$ field theory, and perhaps of SYM on a
non-commuting lower dimensional manifold, lends further support to the
conjecture that these theories exist as consistent quantum mechanical
theories.

\noindent 2. Although we can not localize a state at a region smaller than 
$\zeta$, the theory is still defined in the continuum. If we take a
bounded region in space, of some size much larger than $\zeta$, then
the number of states in this region is infinite. To roughly see how
many states one has in a bounded region in space, we can focus, in the
quantum mechanics, on wave functions that have support in some
bounded, much larger than $\zeta$, region around the origin. In
the resolved sigma model the number of such states is infinite.

\noindent 3. A more interesting point is that we can now also discuss the UV 
of the theory. The simplest case to analyze is the K=1 case. The
theory in the IR contains a free tensor multiplet and we can now probe
the UV by a high momentum exchange scattering process.

For concreteness let us discuss the scattering process of two
particles carrying each one unit of momentum along the null circle
(for the K=1 case, \berdoug). We are therefore interested in the N=2 moduli
space which is $R^4\times R^4/Z_2$. When we resolve the singularity it
is replaced by $R^4\times ALE$. The first component is the center of
mass of the two particles and the second is their relative position. A
2-particle scattering process is described by sending a wave packet on
the ALE from infinity towards the origin. The rough picture will be
that in momentum scales $P$ much smaller than the one set by $\zeta$,
the theory will have an expansion in $\zeta P^2$. At high energy,
however, the wave packet will explore the resolved sphere, which
suggests a qualitative difference in the result, beyond an expansion
by $\zeta P^2$. One also doubts whether the deformed interaction
$e^{\zeta\partial^2}$ captures the large $P$ behavior correctly.

This picture suggests that even though the theory is defined in the
continuum, the UV is not governed by any scale invariance. In fact,
the resolved region of the sigma model, i.e., the one that high
energy processes probe, is the one which is the most sensitive to the
violation of scale invariance by $\zeta$. The principal that
should replace the UV fixed point is not clear.

\vskip 1cm

\centerline{\bf Acknowledgments}\nobreak

We would like to thank O. Aharony, T. Banks, S. Kachru, A. Lawrence,
N. Nekrasov, N. Seiberg, S. Sethi and E. Silverstein. This work is
supported by NSF grant NSF PHY-9512835.

\listrefs

\end